\batchmode
\documentclass{article}
\usepackage[T1]{fontenc}
\usepackage[latin9]{inputenc}
\usepackage{geometry}
\usepackage{amsmath}
\usepackage{graphicx}
\usepackage{amssymb}
\usepackage{esint}

\makeatletter
\newcommand{\lyxaddress}[1]{
\par {\raggedright #1
\vspace{1.4em}
\noindent\par}
}

\newcommand{\dvk}{\mathrm{d}^{4}k}

\newcommand{\nb}{n_\text{B}}
\newcommand{\nf}{n_\text{F}}
\newcommand{\PiT}{\Pi_\text{T}}

\newcommand{\DT}{D_\text{T}}

\newcommand{\I}{\text{Im}}
\newcommand{\R}{\text{Re}}

\newcommand{\GeV}{\text{G}\hspace{-.2px}e\hspace{-.8px}\text{V}}
\newcommand{\MeV}{\text{M}\hspace{-.5px}e\hspace{-.8px}\text{V}}

\usepackage{feynmf}
\usepackage{cite}

\makeatletter\renewcommand{\@fnsymbol}[1]{\ensuremath{%
 \ifcase#1\or \dagger\or **\or {**}*\or   \mathsection\or \mathparagraph\or \|\or \star\or   \star\star\or {\star\star}\star \else\@ctrerr\fi}}\makeatother

\makeatother

\begin{document}

\title{Equation of state for strongly interacting matter: collective effects,
Landau damping and predictions for LHC}

\date{R.~Schulze\thanks{r.schulze@fzd.de, invited talk given at XLVI International Winter Meeting on Nuclear Physics in Bormio, Italy, 20-26 Jan 2008},
M.~Bluhm, B.~Kämpfer}

\maketitle

\lyxaddress{Forschungszentrum Dresden-Rossendorf, PF 510119, 01314 Dresden, Germany\\
Institut f\"ur Theoretische Physik, TU Dresden, 01062 Dresden, Germany}

\begin{abstract}
The equation of state (EOS) is of utmost importance for the description
of the hydrodynamic phase of strongly interacting matter in relativistic
heavy-ion collisions. Lattice QCD can provide useful information on
the EOS, mainly for small net baryon densities. The QCD quasiparticle
model provides a means to map lattice QCD results into regions relevant
for a variety of experiments. We report here on effects of collectives
modes and damping on the EOS. Some predictions for forthcoming heavy-ion
collisions at LHC/ALICE are presented and perspectives for deriving
an EOS for FAIR/CBM are discussed.
\end{abstract}

\section{Introduction}

The equation of state (EOS) for strongly interacting matter is needed
as input for hydrodynamical calculations of the expanding fireball
created in relativistic heavy-ion collisions (HIC). Theoretical predictions
(cf.~\cite{KMR03} for a survey) and recent experimental results
\cite{BRA05,PHO05,STA05,PHE05} point to a transition from confined
hadronic matter to the quark-gluon plasma (QGP), being a new deconfined
state which is governed by the fundamental quark and gluon degrees
of freedom. That means, a usable EOS has to uncover both states.

Upcoming HIC experiments at LHC, mainly to be investigated by ALICE,
will probe the high-temperature region at small net baryon densities,
while the future HIC experiments at FAIR, to be addressed by CBM,
are aimed at exploring the region of high net baryon densities. Therefore,
the EOS in a wide region of the phase diagram is needed. Numerical
simulations of QCD on the lattice are still constrained to small net
baryon densities. Consequently, there is a need for phenomenological
models which allow predictions in regions of the phase diagram not
yet accessible by lattice QCD calculations. Here we discuss a phenomenological
model which relies on the picture of quarks and gluons as non-interacting
quasiparticle excitations. The employed quasiparticle model (QPM)
goes back to \cite{Pes94,Pes96,LH98,Pes00,Pes02}, while recent work
has been presented in \cite{Blu04b,KBS06,BKS06,BKS07a,BKS07b,Sch08}.
Alternative formulations have been given, e.g., in \cite{LR03,TSW04}.

\section{The quasiparticle model}

The description of strongly interacting matter is governed by QCD.
Thus the foundation of any model has to be the quantized Lagrangian
${\cal L}_{\text{QCD}}$ and the dressed propagators and full self-energies
obtained from it. In the framework of finite-temperature field theory
a thermodynamical potential $\Omega$ can be derived using the Cornwall-Jackiw-Tomboulis
(CJT) formalism \cite{CJT74}. It employs the effective action\begin{eqnarray}
\Gamma[D,S] & = & I\,-\,\frac{1}{2}\left\{ \text{Tr}\left[\ln D^{-1}\right]+\text{Tr}\left[D_{0}^{-1}D-1\right]\right\} \nonumber \\
 &  & \,+\,\left\{ \text{Tr}\left[\ln S^{-1}\right]+\text{Tr}\left[S_{0}^{-1}S-1\right]\right\} \,+\,\Gamma_{2}[D,S],\label{eq:cjt effective action}\end{eqnarray}
where $I$ is the classical action containing ${\cal L}_{\text{QCD}}$,
and $D$ and $S$ are the dressed gluon and quark propagators (the
subscript $0$ denotes the respective free equivalents). The functional
$\Gamma_{2}$ represents the sum over all two-particle irreducible
skeleton graphs of the theory, i.e.~all those graphs without external
lines that do not fall apart upon cutting of two propagators.

For translationally invariant systems without broken symmetries the
expression (\ref{eq:cjt effective action}) simplifies and gives the
thermodynamic potential at finite temperature $T$ \cite{Sch08}\begin{eqnarray}
\frac{\Omega}{V} & = & \mbox{tr}\!\!\int\!\!\frac{\dvk}{(2\pi)^{4}}\nb(\omega)\,\mbox{Im}\!\left(\ln D^{-1}-\Pi D\right)\label{eq:Omegafinal}\\
 &  & +\,2\,\mbox{tr}\!\!\int\!\!\frac{\dvk}{(2\pi)^{4}}\nf(\omega)\,\mbox{Im}\!\left(\ln S^{-1}-\Sigma S\right)-\frac{T}{V}\Gamma_{2},\nonumber \end{eqnarray}
where $\Pi$ and $\Sigma$ are the full self-energies of gluons and
quarks respectively. Truncating $\Gamma_{2}$ at 2-loop order leaving
\vspace{.2cm}
\begin{fmffile}{phis}
\begin{equation}
  \vspace{.5cm}
  \Gamma_2 = \frac{1}{12} \,\,\,\parbox{25mm}{\begin{fmfgraph}(39,25) \fmfleft{i}\fmfright{o} \fmf{phantom,tension=1.9}{i,v1}
      \fmf{phantom,tension=0.13}{v1,v2} \fmf{phantom,tension=0.4}{v2,o} \fmffreeze 
      \fmf{photon,right,tension=4}{v1,v2,v1} \fmf{photon,tension=1}{v1,v2} 
      \fmfdot{v1,v2} \end{fmfgraph}} \hspace{-1.3cm} + \,\frac{1}{8} \,\,\,
  \parbox{25mm}{\begin{fmfgraph}(55,25) \fmfleft{i}\fmfright{o} \fmf{phantom,tension=1.8}{i,v1}
      \fmf{phantom,tension=0.08}{v1,v2} \fmf{phantom,tension=0.08}{v2,v3} 
      \fmf{phantom,tension=0.8}{v3,o} \fmffreeze \fmf{photon,right,tension=1}{v2,v1,v2} \fmf{photon,right,tension=1}{v2,v3,v2} 
      \fmfdot{v2} \end{fmfgraph}} 
  \hspace{-5mm} - \,\,\frac{1}{2} \hspace{-1mm}
  \parbox{20mm}{\begin{fmfgraph}(55,25) \fmfleft{i}\fmfright{o} \fmf{phantom,tension=0.4}{i,v1}
      \fmf{phantom,tension=0.13}{v1,v2} \fmf{phantom,tension=0.4}{v2,o} \fmffreeze 
      \fmf{vanilla,right,tension=0.6}{v1,v2,v1} \fmf{photon,tension=1}{v1,v2} 
      \fmfdot{v1,v2} \end{fmfgraph}} \hspace{-3mm}
\end{equation}
\end{fmffile}directly leads to the well-known 1-loop quark and gluon self-energies
\cite{LeB96}. Assuming additionally small external momenta or, equivalently,
hard thermal loops ensures gauge invariance. These approximations
are used in what follows.

An important quantity of the strong interaction and consequently also
of our model is the running coupling $g^{2}$ which depends on the
ratio of renormalization scale and QCD scale parameters. In order
to phenomenologically accommodate higher-order and even non-perturbative
effects of QCD we replace the former at $\mu=0$ ($\mu$ is the quark
chemical potential) by the first Matsubara frequency $i\pi T$ and
shift the temperature $T$ by a parameter $T_{s}$. The new quantity
corresponds to an effective coupling and is denoted by $G^{2}$ (see
\cite{BKS07a} for details):\begin{equation}
G^{2}(T\geq T_{c},\mu=0)=\frac{16\pi^{2}}{\beta_{0}\ln x^{2}}\left(1-\frac{4\beta_{1}}{\beta_{0}^{2}}\frac{\ln\left[\ln x^{2}\right]}{\ln x^{2}}\right),\label{eq:eff coupling}\end{equation}
where $x\equiv\lambda(T-T_{s})/T_{c}$.

From the resulting thermodynamic potential we find the entropy density
$s:=-V^{-1}\left.\partial\Omega/\partial T\right|_{\mu}=s_{g,\text{T}}+s_{g,\text{L}}+\sum_{q,s}(s_{i,\text{Pt.}+}+s_{i,\text{Pl.}})$
as a sum of four seemingly non-interacting quasiparticle families.
Transverse gluons ($g$,T) and the quark particle ($i$,Pt.) contributions
have a real particle interpretation while longitudinal gluons (plasmons,
$g$,L) and the quark plasmino ($i$,Pl.) contributions are collective
modes similar to phonons in solid state physics.

Although no obvious interaction terms appear within the one-loop entropy
density, quark-gluon interactions are incorporated via damping terms
as part of the single quasiparticle contributions. For instance, the
transverse gluon entropy density reads\begin{eqnarray}
s_{g,\text{T}} & = & +2d_{g}\int_{\dvk}\frac{\partial\nb}{\partial T}\Big\{\pi\varepsilon(\omega)\Theta\!\left(-\R\DT^{-1}\right)-\arctan\frac{\I\PiT}{\R\DT^{\text{-}1}}+\mbox{Re}\DT\mbox{Im}\PiT\Big\},\end{eqnarray}
where the first term represents the real quasiparticle entropy density
being equal to the entropy density of an ideal gas but with an implicit
dispersion relation $\omega^{2}=k^{2}+\Pi_{i}(\omega,k)$, where $\Pi_{i}(\omega,k)$
are the respective self-energies of species $i$ which depend furthermore
on temperature and chemical potential. The second and third terms
represent the Landau damping.

At zero chemical potential $\mu$, both damping terms and the plasmon
and plasmino entropies give only small contributions. Omitting those,
an effective quasiparticle model (eQP) can be formulated to simplify
the description/prediction of experiments with negligibly small net
baryon density. This is a good approximation, e.g., for Au+Au collisions
at RHIC or Pb+Pb collisions at LHC. Also note that the eQP uses the
asymptotic approximation of the dispersion relation near the light
cone, $\omega^{2}=k^{2}+m_{i,\infty}^{2}$, with $m_{i,\infty}$ as
asymptotic quasiparticle masses which depend on $T$ and $\mu$ both
explicitly and implicitly (via $G^{2}$).%
\begin{figure}
\noindent \begin{centering}
\includegraphics[bb=0bp 0bp 240bp 183bp,clip,scale=0.9]{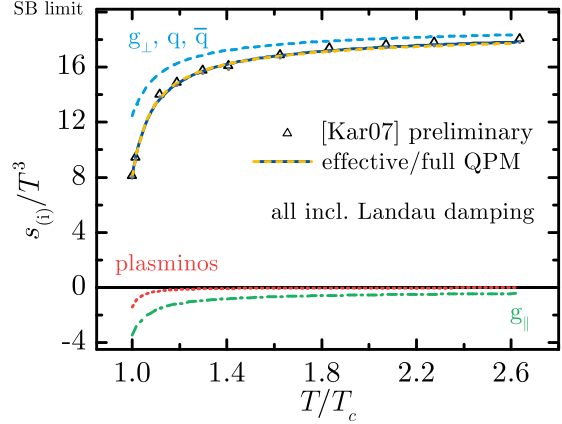}
\par\end{centering}

\caption{Adjustment of both full (solid line, $T_{s}=-0.73T_{c}$ and $\lambda=6.1$)
and effective (dashes on solid line, $T_{s}=-0.75T_{c}$ and $\lambda=6.3$)
QPMs to lattice data for the entropy density $s/T^{3}$ from \cite{Kar07}
(continuum extrapolated by a factor of 0.96). Contributions to the
full QPM are given by the dashed (transverse gluons, quarks, antiquarks),
dotted (plasminos) and dash-dotted (plasmons) lines.\label{fig:entropy at mu eq 0}}

\end{figure}
\begin{figure}[t]
\noindent \begin{centering}
\includegraphics{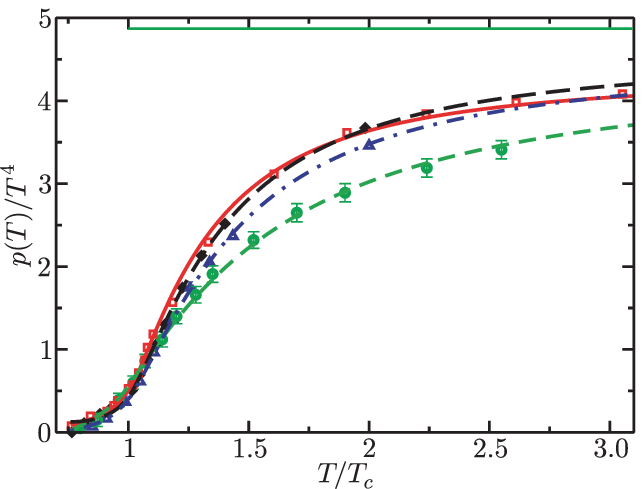}\includegraphics{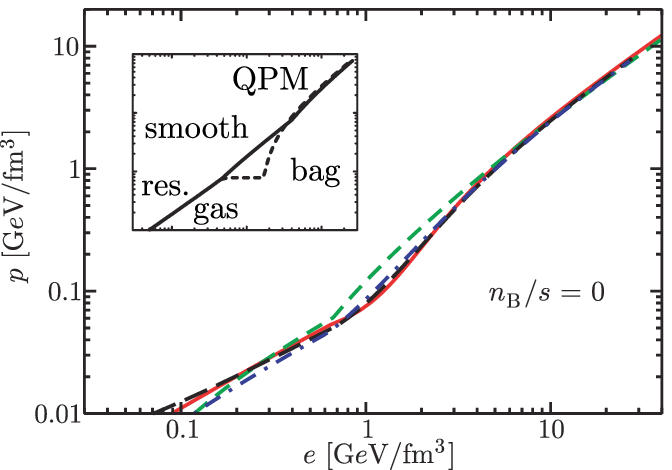}
\par\end{centering}

\caption{Left panel: Adjustments of the scaled eQP pressure $p/T^{4}$ to various
lattice calculations (Ref.~\cite{Kar03} - squares, Ref.~\cite{Ber05}
- diamonds and triangles, Ref.~\cite{Aok06} - circles). Right panel:
Corresponding EOS in the form of pressure $p$ as a function of energy
density $e$. For details see \cite{BKS07b}.\label{fig:eQP EOS}}

\end{figure}

\section{EOS for $\mu\approx0$}

In order to obtain sensible predictions, the QPM is adjusted lattice
data. Exploring the flexibility of the QPM introduced by the effective
coupling we find excellent agreement of both the full QPM and the
effective QPM with lattice data, see Figs.~\ref{fig:entropy at mu eq 0}
and \ref{fig:eQP EOS}. In light of the substantially more involved
nature of the full QPM it is very notable that both models give essentially
the same good description of available lattice data. Fig.~\ref{fig:entropy at mu eq 0}
also shows that plasmons and plasminos give negative contributions
to the entropy density which is due to correlations introduced into
the quark-gluon system by those collective modes.

It is remarkable that even though there still are substantial difference
between various lattice QCD results, which are e.g.~due to the used
actions, lattice sizes and continuum extrapolations, the QPM EOS in
the form $p(e)$ is unique above a threshold of about $e\gtrsim4\,\GeV/\text{fm}^{3}$
(Fig.~\ref{fig:eQP EOS}). Some uncertainty is seen in the regions
of lower energy densities. We suppose that the hadron resonance gas
is the correct description of strongly interacting matter in the confined
region. In order to examine the impact of this uncertainty we investigate
two extreme cases: (i) a smooth crossover (labeled {}``QPM 4.0''),
and (ii) a first order transition between resonance gas and the confident
region of our QPM (labeled {}``bag model'').

\begin{figure}
\noindent \begin{centering}
\includegraphics{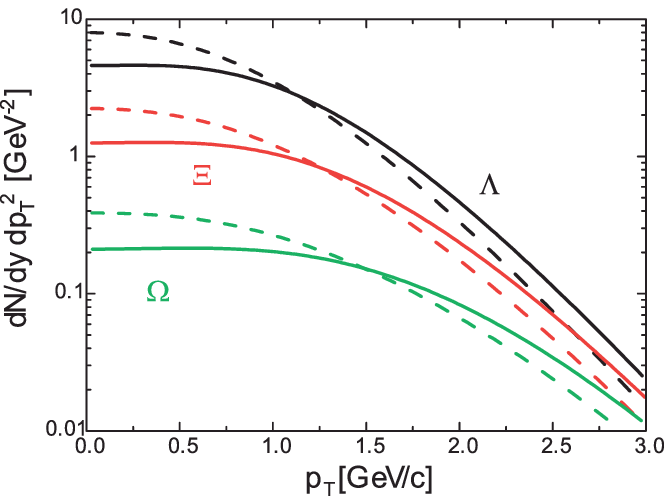}\includegraphics{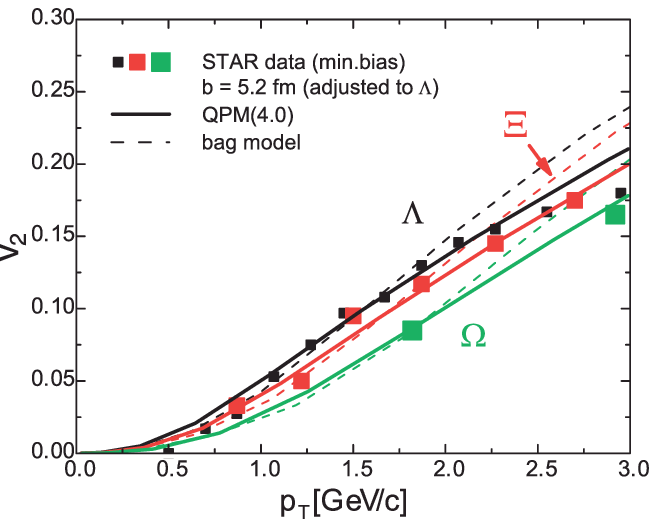}
\par\end{centering}

\caption{Transverse momentum spectra (left) and azimuthal distribution $v_{2}$
of emitted hadrons (right) for some strange baryons. Symbols represent
experimental data \cite{STA05b} for Au+Au collisions at RHIC. For
details see \cite{BKS07b}.\label{fig:RHIC}}

\end{figure}

To do so we use a relativistic hydrodynamic model to simulate Au+Au
collisions at RHIC energies. Fig.~\ref{fig:RHIC} shows the resulting
transverse momentum spectra and the azimuthal anisotropy coefficient
$v_{2}$ of the baryons $\Lambda$, $\Xi$ and $\Omega$ for an initial
state characterized by $s_{0}=110\,\text{fm}^{-3}$ and initial proper
time $\tau_{0}=0.6\,\text{fm}/c$. The latter is compared to actual
experimental results \cite{STA05b}, showing good agreement of both
crossover and first order phase transition in the transverse momentum
region $p_{\text{T}}\lesssim1.8\,\GeV$ considered relevant for hydrodynamics.
Above this region, a simple crossover from the resonance gas to the
QPM clearly provides better description of the measured data.%
\begin{figure}
\noindent \begin{centering}
\includegraphics{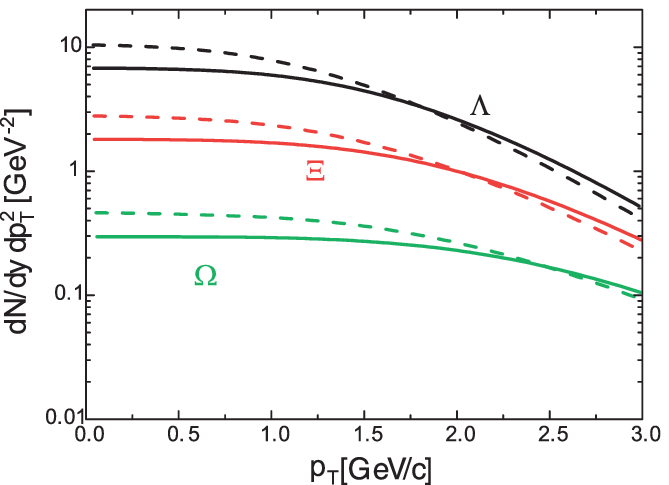}\includegraphics{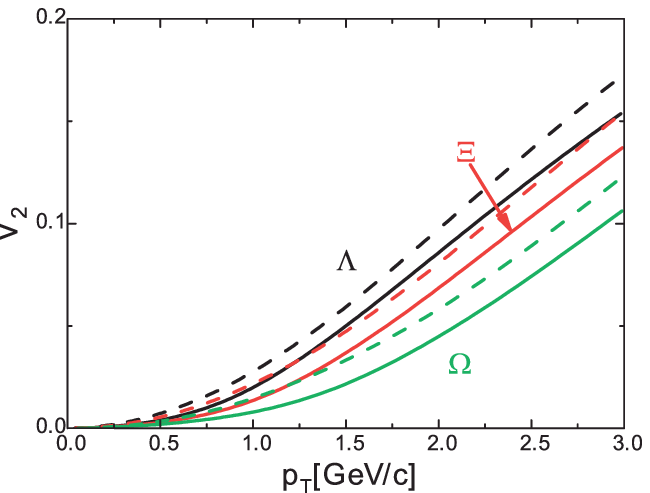}
\par\end{centering}

\caption{Transverse momentum spectra (left) and azimuthal distribution of emitted
hadrons $v_{2}$ (right) for the same particles as in Figure \ref{fig:RHIC}
as predicted for Pb+Pb collisions at LHC. For details see \cite{BKS07b}.\label{fig:LHC}}

\end{figure}

To consider Pb+Pb collisions at LHC, a conservative guess can give
first indications of possible differences to RHIC. LHC particle yields
are assumed to be three times larger than at RHIC, hinting to $s_{0}^{\text{LHC}}=3s_{0}^{\text{RHIC}}=330\,\text{fm}^{-3}$
($T_{0}=515\,\MeV$) with the assumption of $\tau=0.6\,\text{fm}/c$.
The higher initial temperature at LHC leading to a longer fireball
lifetime suggests a stronger radial flow as well as a more equilibrated
azimuthal distribution of emitted hadrons. Indeed, the predicted $p_{\text{T}}$
spectrum for $\Lambda$, $\Xi$ and $\Omega$ is considerably flatter
than at RHIC, while $v_{2}$ is noticeably reduced (Fig.~\ref{fig:LHC}).

In these examples the effects of a non-zero baryon density are negligibly
small.

\section{Nonzero net baryon density}

The advantage of the phenomenological QPM is its ability to provide
an EOS at nonzero chemical potential, in particular in a region which
is expect to be relevant for forthcoming experiments at FAIR. This
remarkable ability is due to the thermodynamic self-consistency of
the QPM. As a consequence, thermodynamic quantities at arbitrary values
of the state variables (here $\mu$ and $T$) are connected through
Maxwell relations and the stationarity condition of the thermodynamic
potential. Thus the model is able to map the lattice data at $\mu=0$
into the $T$-$\mu$-plane. This is achieved by solving the Maxwell
relation, which is a partial differential equation of first order
for the effective coupling $G^{2}(T,\mu)$, using the method of characteristics
with the parametrized $G^{2}(T,\mu=0)$ as initial condition. This
procedure has been tested successfully against lattice calculations
of the pressure corrections coefficients available for small chemical
potential \cite{Blu04b,BKS07b}.

However, the eQP, where damping terms and collective modes are neglected,
meets some ambiguity in the region of large chemical potential and
not too high temperatures, since the characteristic curves of the
partial differential equation exhibit crossings. Consequently, for
mapping to large net baryon densities the full QPM has to be applied.
As a sign of the self-consistency of the latter one, no crossings
appear among its characteristics \cite{Sch08}. Thus, the effective
coupling $G^{2}$ is unique for every point of the $T$-$\mu$-plane.
From the effective coupling $G^{2}(T,\mu)$ entropy density $s$ and
net quark number density $n$ follow directly as thermodynamic integrals.
However, to ensure the physical relevance of the solutions, agreement
with general thermodynamic requirements, e.g.~Nernst's theorem, has
to be verified. %
\begin{figure}[t]
\begin{centering}
\includegraphics[bb=40bp 360bp 260bp 550bp,clip]{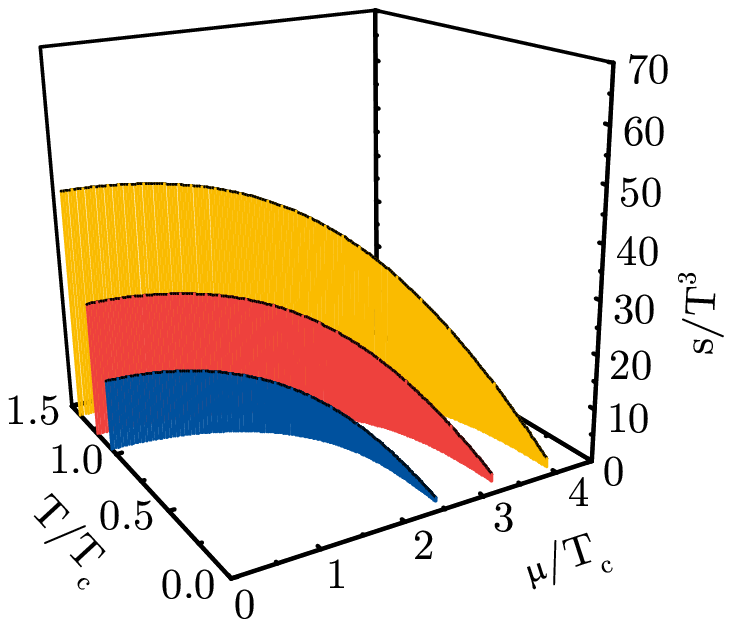}\includegraphics[bb=40bp 360bp 260bp 550bp,clip]{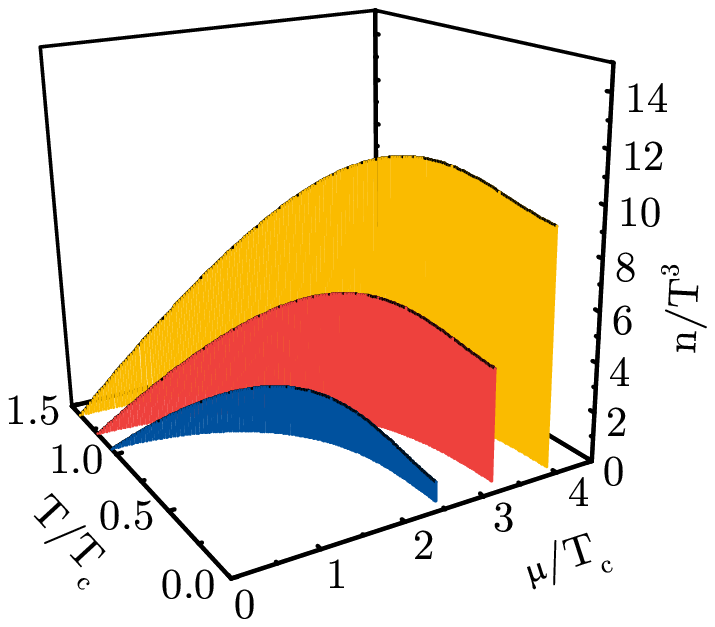}
\par\end{centering}

\caption{Scaled entropy density $s/T^{3}$ (left) and scaled net baryon density
$n/T^{3}$ (right) of strongly interacting matter as predicted by
the full QPM. Both quantities are exhibited along selected characteristic
curves.\label{fig:HTL s and n}}

\end{figure}
\begin{figure}[!t]
\begin{centering}
\includegraphics[bb=0bp 0bp 280bp 195bp,clip,scale=0.8]{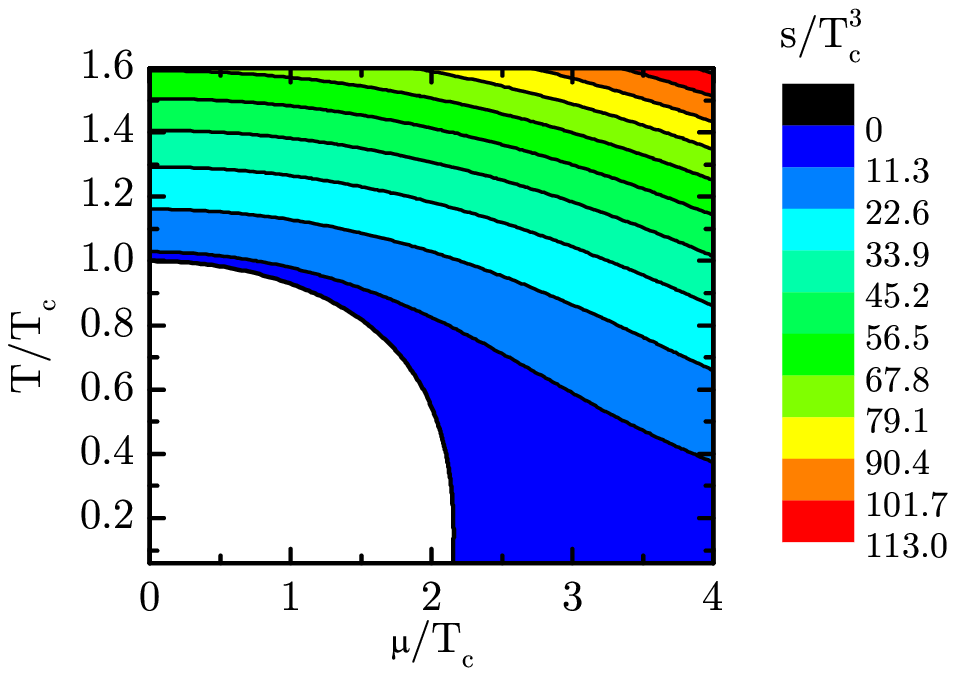}\includegraphics[bb=0bp 0bp 280bp 195bp,clip,scale=0.8]{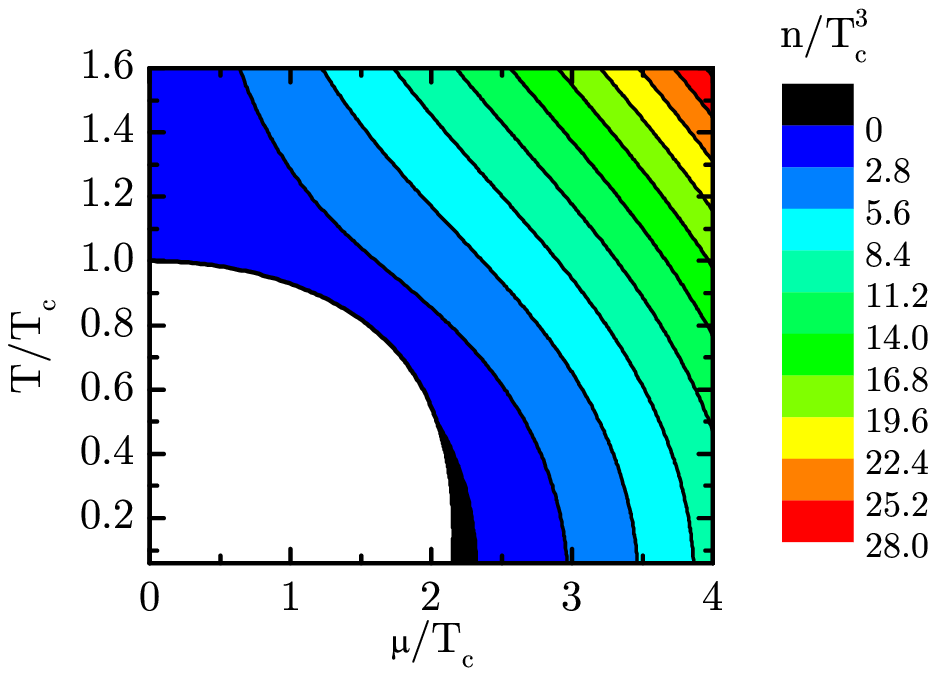}
\par\end{centering}

\caption{Contour plot of the scaled entropy density $s/T^{3}$ (left) and scaled
net baryon density $n/T^{3}$ (right) as a function of temperature
$T$ and chemical potential $\mu$ from Fig.~\ref{fig:HTL s and n}.\label{fig:HTL s contour}}

\end{figure}

Indeed, Fig. \ref{fig:HTL s and n}, which shows both quantities along
selected characteristics starting above $T_{c}$, confirms that the
entropy density vanishes for $T\rightarrow0$ and the net number density
increases with the chemical potential. Contour plots of the thermodynamic
quantities (Fig.~\ref{fig:HTL s contour}) also show regular behavior
of the thermodynamic quantities for the region above the expected
phase transition. Therefore, the full QPM can be used to predict a
physical EOS for strongly interacting matter especially at high net
baryon densities.

Below the {}``phase transition'', the model, in the present version,
cannot directly be applied since strong contributions of collective
modes lead to a negative baryon density. It remains to be checked
whether improved dispersion relations and a refined treatment of the
imaginary parts of the self-energies can cure this obstacle. However,
for most cases it is more prudent to use the resonance gas below the
phase transition as shown for the eQP, so that these ambiguities do
not pose a serious problem. The resulting {}``compound EOS'' can
then not only be used for predictions of upcoming experiments at CBM@FAIR
but also as an input to general relativistic models of compact stellar
objects such as neutron/quark/strange stars.

\section{Conclusion}

Our quasiparticle model in both the previous simplified version and
the extended version with collective modes and Landau damping is able
to simultaneously describe recent lattice calculations at zero and
small chemical potential. Employing the resulting equation of state,
combined with a resonance gas model, in a hydrodynamical code the
experimental data from RHIC are fairly well described. Furthermore,
predictions for heavy-ion experiments at LHC can be given. For both
experimental situations the simplified, effective quasiparticle model
suffices due to small net baryon densities. However, for larger net
baryon densities, the full model including the suitably parametrized
HTL dispersion relations, Landau damping and collective modes has
to be employed. The current results are encouraging with respect of
deriving an equation of state usable in a large region of the phase
diagram of strongly interacting matter.

~\\
\emph{Acknowledgment}: R.S.~would like to thank the organizers for
the invitation to this very insightful and inspiring meeting and the
financial support granted.


\begin{thebibliography}{25}

\bibitem{KMR03}
J.~Kapusta, B.~M\"uller, J.~Rafelski, \emph{Quark-Gluon Plasma: Theoretical
  Foundations} (Elsevier, 2003), ISBN 0444511105

\bibitem{BRA05}
I.~Arsene et~al. (BRAHMS), Nucl. Phys. A \textbf{757}, 1 (2005),
  \texttt{nucl-ex/0410020}

\bibitem{PHO05}
B.B. Back et~al. (PHOBOS), Nucl. Phys. A \textbf{757}, 28 (2005),
  \texttt{nucl-ex/0410022}

\bibitem{STA05}
J.~Adams et~al. (STAR), Nucl. Phys. A \textbf{757}, 102 (2005),
  \texttt{nucl-ex/0501009}

\bibitem{PHE05}
K.~Adcox et~al. (PHENIX), Nucl. Phys. A \textbf{757}, 184 (2005),
  \texttt{nucl-ex/0410003}

\bibitem{Pes94}
A.~Peshier, B.~K\"ampfer, O.P. Pavlenko, G.~Soff, Phys. Lett. B \textbf{337},
  235 (1994)

\bibitem{Pes96}
A.~Peshier, B.~K\"ampfer, O.P. Pavlenko, G.~Soff, Phys. Rev. D \textbf{54}(3),
  2399 (1996)

\bibitem{LH98}
P.~L\'evai, U.~Heinz, Phys. Rev. C \textbf{57}, 1879 (1998),
  \texttt{hep-ph/9710463}

\bibitem{Pes00}
A.~Peshier, B.~K\"ampfer, G.~Soff, Phys. Rev. C \textbf{61}, 045203 (2000),
  \texttt{hep-ph/9911474}

\bibitem{Pes02}
A.~Peshier, B.~K\"ampfer, G.~Soff, Phys. Rev. D \textbf{66}, 094003 (2002),
  \texttt{hep-ph/0206229}

\bibitem{Blu04b}
M.~Bluhm, B.~K\"ampfer, G.~Soff, Phys. Lett. B \textbf{620}, 131 (2005),
  \texttt{hep-ph/0411106}

\bibitem{KBS06}
B.~K\"ampfer, M.~Bluhm, R.~Schulze, D.~Seipt, U.~Heinz, Nucl. Phys. A
  \textbf{774}, 757 (2006), \texttt{hep-ph/0509146}

\bibitem{BKS06}
M.~Bluhm, B.~K\"ampfer, R.~Schulze, D.~Seipt, Acta Phys. Hung. A \textbf{27},
  397 (2006), \texttt{hep-ph/0608052}

\bibitem{BKS07a}
M.~Bluhm, B.~K\"ampfer, R.~Schulze, D.~Seipt, Eur. Phys. J. C \textbf{49}, 205
  (2007), \texttt{hep-ph/0608053}

\bibitem{BKS07b}
M.~Bluhm, B.~K\"ampfer, R.~Schulze, D.~Seipt, U.~Heinz, Phys. Rev. C
  \textbf{76}, 034901 (2007), \texttt{arXiv:0705.0397}

\bibitem{Sch08}
R.~Schulze, M.~Bluhm, B.~K\"ampfer, Eur. Phys. J. ST  (2008), in print,
  \texttt{arXiv:0709.2262}

\bibitem{LR03}
J.~Letessier, J.~Rafelski, Phys. Rev. C \textbf{67}, 031902 (2003),
  \texttt{hep-ph/0301099}

\bibitem{TSW04}
M.A. Thaler, R.A. Schneider, W.~Weise, Phys. Rev. C \textbf{69}, 035210 (2004),
  \texttt{hep-ph/0310251}

\bibitem{CJT74}
J.M. Cornwall, R.~Jackiw, E.~Tomboulis, Phys. Rev. D \textbf{10}, 2428 (1974)

\bibitem{LeB96}
M.L. Bellac, \emph{Thermal Field Theory} (Cambridge University Press, 1996),
  ISBN 0521654777

\bibitem{Kar07}
F.~Karsch (2007), \texttt{hep-ph/0701210}

\bibitem{Kar03}
F.~Karsch, K.~Redlich, A.~Tawfik, Phys. Lett. B \textbf{571}, 67 (2003),
  \texttt{hep-ph/0306208}

\bibitem{Ber05}
C.~Bernard et~al., PoS \textbf{LAT2005}, 156 (2006), \texttt{hep-lat/0509053}

\bibitem{Aok06}
Y.~Aoki, Z.~Fodor, S.D. Katz, K.K. Szab\'{o}, JHEP \textbf{2006}, 089 (2006),
  \texttt{hep-lat/0510084}

\bibitem{STA05b}
J.~Adams et~al. (STAR Collaboration), Phys. Rev. Lett. \textbf{95}, 152301
  (2005), \texttt{nucl-ex/0501016}

\end{thebibliography}
\end{document}